\documentclass[prl,twocolumn,twoside]{revtex4}

\usepackage{graphicx}
\usepackage{amsmath,bm}
\usepackage{ae}

\newcommand{\eq}[1]{\begin{equation}#1\end{equation}}
\newcommand{\eqmulti}[1]{\begin{equation}\begin{split}#1\end{split}\end{equation}}

\newcommand{\ket}[1]{\ensuremath{\,|{#1}\rangle}}

\newcommand{\matrixe}[3]{\ensuremath{\langle{#1}|\,{#2}\,|{#3}\rangle}}

\newcommand{\op}[1]{\ensuremath{\bm{#1}}}

\newcommand{\aO}{\ensuremath{\op{a}}}
\newcommand{\aaO}{\ensuremath{\op{a}^{\dag}}}
\newcommand{\nO}{\ensuremath{\op{n}}}

\newcommand{\HO}{\ensuremath{\op{H}}}

\begin{document}

\title{Ultracold Bosonic Atoms in Disordered Optical Superlattices}

\author{R. Roth}
\email{robert.roth@physics.ox.ac.uk}

\author{K. Burnett}
\email{k.burnett1@physics.ox.ac.uk}

\affiliation{Clarendon Laboratory, University of Oxford,
  Parks Road, Oxford OX1 3PU, United Kingdom}

\date{\today}

\begin{abstract}   
The influence of disorder on ultracold atomic Bose
gases in quasiperiodic optical lattices is discussed in the framework of
the one-dimensional Bose-Hubbard model. It is shown that simple periodic
modulations of the well depths generate a rich phase diagram consisting of
superfluid, Mott insulator, Bose-glass and Anderson localized phases. The
detailed evolution of mean occupation numbers and number fluctuations as
function of modulation amplitude and interaction strength is discussed.
Finally, the signatures of the different phases, especially of the
Bose-glass phase, in matter-wave interference experiments are
investigated.
\end{abstract}

\maketitle


Quantum phase transitions of interacting bosonic many-body systems  in
disordered lattice potentials have been a topic of intense theoretical
investigations in the past years. A rich variety of possible
zero-temperature phases has been predicted such as the Mott-insulator
phase, a Anderson localized phase, and a Bose glass phase \cite{FiWe89,
ScBa91, KrTr91, FrMo96}. Recently, impressive experiments on the
transition from a superfluid to a Mott-insulator \cite{GrMa02}
demonstrated that ultracold bosonic atoms in optical lattices offer unique
possibilities to explore the phase diagram of these systems. The degree of
experimental control is remarkable: The geometry of the lattice potential
can be designed and specified precisely, even disorder can be introduced
in a controlled manner \cite{GuTr97,JaBr98}. The strength of the two-body
interaction can be chosen by means of Feshbach resonances. Ultimately, the
structure of the ground state can be examined in detail. This makes them a
promising candidate to study the competition between disorder and
interaction experimentally in an unprecedented way.

In this Letter we discuss the ground state phase diagram of an interacting
bosonic many-body system in an optical superlattice, i.e., a quasiperiodic
lattice composed of disordered unit cells. We study the dependence of the
ground state on the interaction strength and the amount of disorder within
the Bose-Hubbard model.


Consider a gas of $N$ bosonic atoms in an one-dimensional lattice
potential at zero temperature. For a sufficiently strong lattice it is
convenient to describe the state of the system in a basis of localized
``tight-binding'' wave functions, i.e., the Wannier functions that result
from a band-structure calculation. We assume that only the localized
ground state of each lattice well contributes and excited vibrational
states can be neglected. For a system with $I$ lattice sites the many-body
state can thus be represented in terms of number states 
$\ket{n_1,...,n_I}$ with occupation numbers $n_i$ for the individual sites
$i=1,...,I$. A complete basis of the model space is formed by the set of
number states $\ket{n_1^{\alpha},...,n_I^{\alpha}}$ ($\alpha=1,...,D$)
with all compositions of the occupation numbers. The dimension $D =
(N+I-1)!/[N!\,(I-1)!]$ of the number basis grows dramatically; for $I=8$
lattice sites and $N=8$ particles the dimension is $6435$, for $I=N=10$ it
is already $92378$. Within the model space any state can be expanded in
this number basis
\eq{ \label{eq:bh_state}
  \ket{\psi} 
  = \sum_{\alpha=1}^{D} C_{\alpha}\; \ket{n_1^{\alpha},...,n_I^{\alpha}}
}
with expansion coefficients $C_{\alpha}$. We introduce creation operators
$\aO_i$ and annihilation operators $\aaO_i$, which create and annihilate,
resp., a particle in the lowest vibrational state at site $i$, and the
corresponding occupation number operators $\nO_i=\aaO_i\aO_i$. 

The Hamiltonian of the interacting many-body system in second
quantization is the so-called Bose-Hubbard Hamiltonian \cite{FiWe89,
ScBa91, KrTr91, FrMo96}
\eqmulti{ \label{eq:bh_hamiltonian}
  \HO 
  = \sum_{i=1}^{I} \Big[ 
   &-J\; (\aaO_i \aO_{i+1} + \aaO_{i+1} \aO_{i}) \\
   &+ \epsilon_i\; \nO_i + \frac{V}{2}\; \nO_i (\nO_i-1)  \Big] . 
}
The first term describes the coupling between neighboring sites with a
strength $J$. We use cyclic boundary conditions, i.e. hopping between the
first and the last site of the lattice is included. The last two terms of
\eqref{eq:bh_hamiltonian} give the on-site single-particle energy
$\epsilon_i$ and the on-site two-body interaction with a strength $V$.
Formally, the parameters are given by matrix elements of components of the
coordinate space Hamiltonian calculated with the Wannier wave functions
associated with the individual sites. The hopping strength $J$, e.g., is
the off-diagonal matrix element of the kinetic energy operator calculated
in the Wannier basis. For an irregular lattice potential the on-site
energies $\epsilon_i$ depend explicitly on the site index $i$; for
simplicity we neglect the site-dependence of $J$ and $V$. Interactions
between particles at different sites and long range hopping are also
neglected.

To determine the ground state of the system we solve the eigenvalue
problem of the Bose-Hubbard Hamiltonian numerically. The Hamilton matrix
in the number basis is easy to calculate; the on-site energy and the
two-body interaction form the diagonal and the hopping term generates a
few off-diagonal matrix elements. Since the Hamilton matrix is very sparse
and we are interested in the lowest eigenstates only, an iterative
Lanczos-type algorithm is most efficient to solve the eigenvalue problem. 
This enables us to treat systems with up to $I=12$ and $N=12$ on a
standard PC without further approximations. Larger systems can be treated
by Monte Carlo techniques \cite{BaRo02,BaSc92}. Two relevant ground state
observables we consider in the following are the mean occupation number 
$\bar{n}_i = \matrixe{\psi}{\nO_i}{\psi}$ and number fluctuations
$\sigma_i^2 = \matrixe{\psi}{\nO_i^2}{\psi} -
\matrixe{\psi}{\nO_i}{\psi}^2$ at the individual sites.


The interplay between the three terms of the Bose-Hubbard Hamiltonian
\eqref{eq:bh_hamiltonian} generates a rich zero-temperature phase-diagram.
Its basic structure can be understood by analyzing the contributions of
the different terms to the energy expectation value for particular
states. The off-diagonal hopping term gives a negative contribution to the total
energy if the state is a superposition of many number states. If only the
hopping is present the coefficients of the ground state
\eqref{eq:bh_state} are related to the multinomial coefficients, thus all
number states contribute and number fluctuations $\sigma_i$ are large. The
particles can tunnel freely through the lattice and the system resembles a
superfluid. 

If a repulsive two-body interaction is included then number states with
large occupation numbers at individual sites have high energy expectation
values. The repulsion favors homogeneous distributions of the particles
over all sites. The competition between hopping term and two-body
interaction governs one kind of quantum phase transition present in these
systems. For small interaction strengths $V/J$ the hopping term dominates
and the ground state is a superposition of many number states --- a
superfluid. With increasing $V/J$ those number states with large
occupation numbers at some sites are gradually suppressed because of their
large interaction energy. 

For integer (commensurate) fillings $N/I$ there is a unique number state
with $n_i=N/I$ at each site which minimizes the expectation value of the
two-body interaction. For sufficiently strong interactions the ground
state is given by just this number state. Therefore number fluctuations
$\sigma_i$ and also the expectation value of the hopping term vanish. 
This phase in which tunneling of the particles is inhibited by the
repulsive interaction is called Mott insulator phase.  Quantum Monte Carlo
calculations for the infinite one-dimensional Bose-Hubbard model with
$N/I=1$ show that the superfluid to Mott-insulator transition occurs at
$(V/J)_{\text{crit}}\approx 4.65$ \cite{BaSc92}, which is confirmed by a
renormalization group study \cite{SiRo92} and a strong-coupling expansion
\cite{FrMo96}. Mean-field models predict a much larger critical
interaction strength \cite{KrCa92,JaBr98}.



\begin{figure}
  \includegraphics[width=0.4\columnwidth]{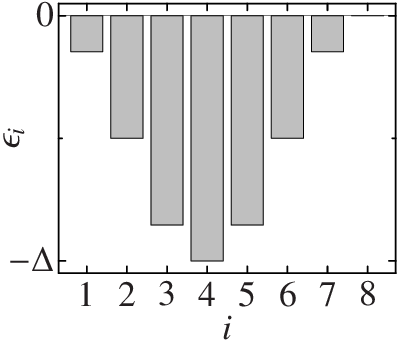}
  \hspace{0.05\columnwidth}
  \parbox[b]{0.4\columnwidth}{
    \caption{On-site energies $\epsilon_i$ for a unit cell 
       of the sinusoidal superlattice.}
    \label{fig:topo_sinusoid08}
    \vspace{3ex} 
  }
\end{figure}

How does disorder affect the phase diagram of the Bose-Hubbard model? 
Most of the theoretical investigations on disorder induced effects
concentrate on infinite lattices with completely random on-site energies.
More relevant for cold bosonic atoms in optical lattices are quasiperiodic
structures, i.e., lattices composed of disordered unit cells.  As simplest
example we discuss an one-dimensional optical superlattice with a
sinusoidal modulation of the on-site energies $\epsilon_i$. A unit cell 
consists of $I=8$ lattice sites and the $\epsilon_i$ vary in the interval
$[-\Delta,0]$ as shown in Fig. \ref{fig:topo_sinusoid08}. Experimentally
this modulation can be realized using a superposition of two optical
standing waves with appropriate wavelengths --- a so-called two-color
lattice.  This allows the independent control of the tunneling strength
$J$ and the disorder amplitude $\Delta$. The sinusoidal modulation already
shows the relevant fundamental features but is only the most simple
realization of disorder, more complex superlattices were already generated
experimentally \cite{GuTr97}.

Let us start from a noninteracting superfluid in a regular lattice
($\Delta/J=0$, $V/J=0$) and increases the disorder amplitude $\Delta$
gradually. The mean occupation number $\bar{n}_i$ of the site with lowest
$\epsilon_i$ in each unit cell will increase. For sufficiently strong
disorder the ground state will be composed of number states which have
non-zero occupation only at the deepest well of each unit cell. This
mechanism is similar to Anderson localization in an infinite system with
completely random on-site energies \cite{BeKi94} and we will adopt the
name for simplicity. However, in contrast to the Anderson localized phase
in a random lattice the ground state is still a superposition of several
number states with considerable number fluctuations $\sigma_i$. 
Therefore, this state appears almost like a superfluid state in a regular
lattice with $I$ times the lattice spacing.

\begin{figure}
  \includegraphics[width=0.79\columnwidth]{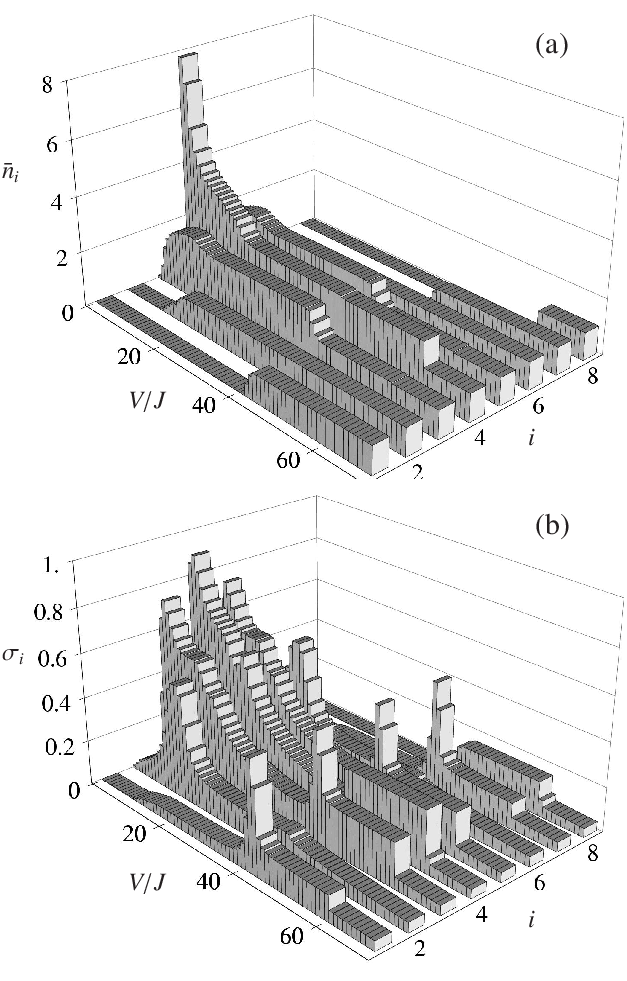}
  \caption{Mean occupation numbers $\bar{n}_i$ (a) and number fluctuations
    $\sigma_i$ (b) for the sinusoidal superlattice with $I=8$, $N=8$, and
    $\Delta/J=60$ as function of $V/J$. The different rows correspond to the 
    individual sites of a unit cell.}
  \label{fig:config_sinusoid08_NN08}
\end{figure}  

The disorder induced localization is strongly affected by repulsive
interactions which drive the system towards even distributions of
particles over all sites. To study the detailed interplay between
interactions and disorder we diagonalize the Hamiltonian numerically for
an isolated $I=8$ unit cell. This assumes that the exchange of particles
between different unit cells can be neglected. By direct comparison with
two cell calculations this turns out to be a very good approximation
for $V/J>0$. 

Figure \ref{fig:config_sinusoid08_NN08} shows the mean occupation numbers
$\bar{n}_i$ and the number fluctuations $\sigma_i$ for the sinusoidal
superlattice with $\Delta/J=60$ and commensurate filling ($I=8$, $N=8$) as
function of the interaction strength $V/J$. For vanishing interaction
strength all particles are localized at the lowest energy site within the
unit cell. However, with increasing $V/J$ the mean occupation at this site
is reduced rapidly and the particles are redistributed to sites with
higher on-site energies. This rearrangement happens stepwise, i.e., first
the two sites with second lowest $\epsilon_i$ are populated. Then at
$V/J\approx10$ the population of the next two sites increases, and so on.
Between successive rearrangements there are extended plateaus (e.g. for
$20<V/J<40$) with constant and approximately integer mean occupation
numbers $\bar{n}_i$. This characteristic sequence of rearrangements and
stable plateaus is also reflected in the number fluctuations $\sigma_i$
shown in Fig.  \ref{fig:config_sinusoid08_NN08}(b). Within the
rearrangement regions number fluctuations of those sites which change
their mean occupation number are large. Within the plateaus all $\sigma_i$
are small and the ground state is almost a pure number state. This region
of successive rearrangements is called Bose-glass phase
\cite{ScBa91,KrTr91}.

Eventually, at $V\approx\Delta$ a final rearrangement happens and all
sites have equal $\bar{n}_i=1$ and very small number fluctuations. This is
the transition from the Bose glass to the Mott insulator phase where the
ground state is a pure number state with $n_i=1$ at each site.

\begin{figure}
  \includegraphics[width=0.8\columnwidth]{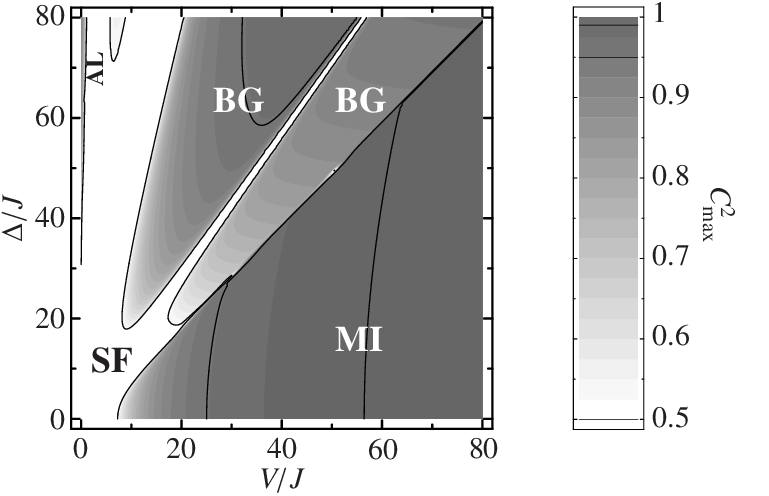}
  \caption{Contour plot of the square of the largest coefficient
    $C_{\max}^2$ for the sinusoidal superlattice with $I=8$, $N=8$ as function
    of $V/J$ and $\Delta/J$. Dark shadings indicate that the ground state
    is a pure number state (insulators).  The labels identify
    the different phases: superfluid (SF), Mott insulator (MI), Bose glass
    (BG), and Anderson localized (AL). }
  \label{fig:phasediag_sinusoid08_NN08}
\end{figure}  

Figure \ref{fig:phasediag_sinusoid08_NN08} summarizes the phase-diagram
for the sinusoidal superlattice. The contour plot shows the square of the
largest coefficient $C_{\max}^2=\max(C_{\alpha}^2)$ in the expansion
\eqref{eq:bh_state}. The dark shadings indicate that the ground state is a
pure number state. The Mott insulator phase (MI) appears only for
$V>\Delta$. For interaction strengths below this boundary two stable
configurations within the Bose-glass phase (BG) follow. At very small
$V/J$ and large disorder amplitudes the Anderson localized phase (AL)
appears. The remaining regions resemble a (disordered) superfluid (SF).

The structure of the Bose glass phase, i.e. the sequence of
rearrangements, depends strongly on the particular pattern of on-site
energies. More complex modulations of the $\epsilon_i$ generate a larger
variety of configurations in the Bose glass phase. The same holds true for
non-commensurate fillings. In the sinusoidal lattice with $N=9$ extended
regions emerge, where two sites exhibit mean occupation $\bar{n}_i=0.5$ or
$\bar{n}_i=1.5$ in association with large number fluctuations.


How can these structures be observed experimentally?  An experimentally
quite simple approach is the imaging of matter-wave interference patterns
after release from the trap and ballistic expansion \cite{GrMa02,OrTu01}. 
The crucial quantity that determines the presence or absence of an
interference pattern is the phase coherence between atoms at different
lattice sites. Small phase fluctuations at the individual sites, which
imply phase coherence, are connected to large number fluctuations
$\sigma_i$. An estimate for the phase fluctuations $\sigma_{\phi,i}$ can
be obtained from the relation $\sigma_{\phi,i}= 1/(2 \sigma_i)$
\cite{OrTu01}. To simulate the density interference pattern for a given
ground state we approximate the wave function of an atom localized at a
lattice site by a Gaussian wave-packet 
\footnote{The centroids of the Gaussians for the different sites have a
constant distance $\Delta\xi$. The width $0.025\Delta\xi^2$ is chosen such
that the typical three-peak structure for the superfluid phase
\cite{GrMa02} is reproduced.}.
The superposition of the wave packets after free expansion (neglecting
interactions) results in a density interference pattern. To account for
the  phase fluctuations we calculate the incoherent average over typically
10000 sets of on-site phases, which are chosen randomly with a Gaussian
distribution of width $\sigma_{\phi,i}$.

\begin{figure}
  \includegraphics[width=0.84\columnwidth]{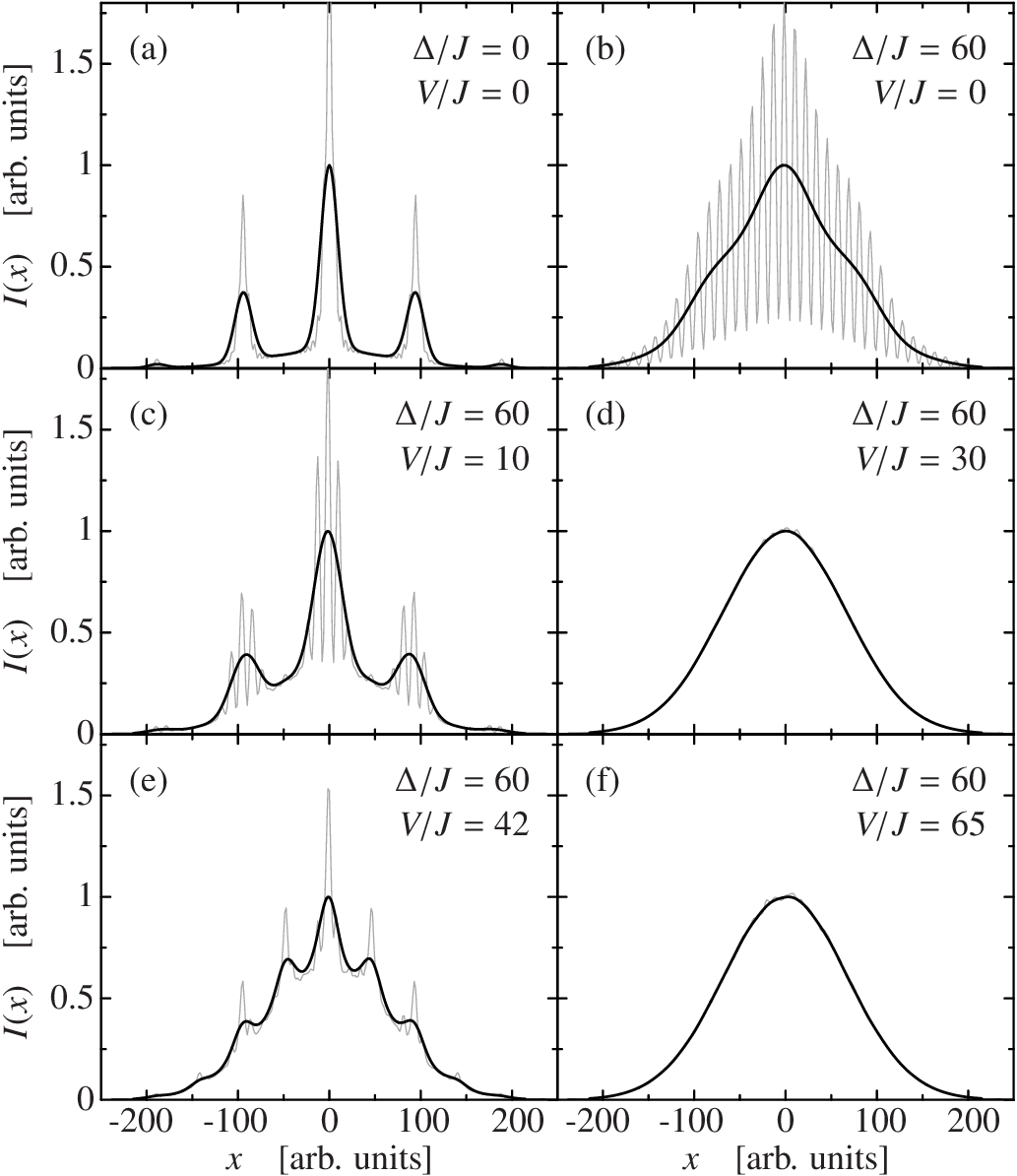}
  \caption{Matter-wave interference patterns for different combinations 
  of $V/J$ and $\Delta/J$. The gray curves show the full interference 
  pattern of the superlattice; the black curves result from 
  folding with a Gaussian profile to mimic a restricted 
  experimental resolution.}
  \label{fig:interf_sinusoid08_NN08}
\end{figure}  

Figure \ref{fig:interf_sinusoid08_NN08} depicts characteristic interference
patterns for different points of the phase diagram. The superlattice
character is explicitly included by replicating the unit cell. This gives
rise to the short wave length oscillations which reflect the periodicity
of the superlattice. The black curves are obtained by folding with a Gaussian
profile to mimic a restricted experimental resolution. The remaining smooth
pattern is largely determined by the structure within the unit cells.

In the absence of interactions and disorder we obtain the prominent
interference pattern of the superfluid \cite{GrMa02} with pronounced peaks
shown in Fig. \ref{fig:interf_sinusoid08_NN08}(a). This structure
dissolves if we enter the Anderson localized phase shown in panel (b) for
$\Delta/J=60$, depending on the experimental resolution one may detect the
interference pattern of a lattice with $I$ times the fundamental lattice
spacing.  If we include interactions $V/J>0$ and enter the Bose glass
phase we observe a characteristic vanishing and reappearance of
interference fringes which is correlated to the number fluctuations shown
in Fig. \ref{fig:config_sinusoid08_NN08}(b). For weak interactions
$V/J=10$  the three-peak structure of the superfluid reappears with
broader peaks and increased incoherent background due to the reduced
number fluctuations. In the stable regions of the Bose glass phase, e.g.
at $V/J=30$, number fluctuations of all sites are small and the fringes
vanish again; only the incoherent background remains as shown in panel
(d). In the vicinity of rearrangements certain lattice sites regain large
number fluctuations and cause the reappearance of interference fringes.
Figure \ref{fig:interf_sinusoid08_NN08}(e) depicts an example for
$V/J=42$, where every second lattice site has considerable fluctuations
(compare Fig. \ref{fig:config_sinusoid08_NN08}) which create a
distinctive fringe pattern. Eventually, if we enter the Mott insulator
phase only the incoherent bump remains as shown in panel (f).
To obtain complementary information on the spatial density distribution 
of these disordered states Bragg diffraction \cite{WeHe95, BiGa95, GuTr97} 
would be a useful tool.


In summary, we have shown that ultracold atomic gases in optical
superlattices are an ideal system to study the complicated interplay
between interaction and disorder. As function of the interaction strength
one can observe the detailed evolution of the ground state from an
Anderson localized state through a Bose-glass to the Mott-insulator, which
is accompanied by a characteristic vanishing and reappearance of
interference fringes. The remarkable degree of experimental access to all
relevant parameters allows a comprehensive study of these quantum phase
transitions.

This work was supported by the Deutsche Forschungsgemeinschaft, the UK
EPSRC, and the EU under the TMR network ERB-FMRX-CT-0002.



\end{document}